\documentclass[fleqn,usenatbib]{mnras}

\usepackage{newtxtext,newtxmath}

\usepackage[T1]{fontenc}
\usepackage{ae,aecompl}

\usepackage{graphicx}	
\usepackage{amsmath}	
\usepackage{amssymb}	

\usepackage{color}

\title[Flares in PSR J1023+0038]{Variable polarisation and Doppler tomography of PSR J1023+0038 - Evidence for the magnetic propeller during flaring?}

\author[P. Hakala and J. J. E. Kajava]{
Pasi Hakala,$^{1}$\thanks{E-mail: pahakala@utu.fi}
Jari J. E. Kajava,$^{1}$
\\
$^{1}$Finnish Centre for Astronomy with ESO (FINCA), University of Turku, V\"{a}is\"{a}l\"{a}ntie 20, FIN-21500 Piikki\"{o}, Finland
}

\date{Accepted XXX. Received YYY; in original form ZZZ}

\pubyear{2015}

\begin{document}
\label{firstpage}
\pagerange{\pageref{firstpage}--\pageref{lastpage}}
\maketitle

\begin{abstract}
Transitional millisecond pulsars are systems that alternate between an accreting low-mass X-ray binary (LMXB) state, and a non-accreting radio pulsar state.
When at the LMXB state, their X-ray and optical light curves show rapid flares and dips, origin of which is not well understood.
We present results from our optical and NIR observing campaign of PSR J1023+0038, a transitional millisecond pulsar observed in an accretion state. 
Our wide band optical photopolarimetry indicates that the system shows intrinsic linear polarisation, the degree of which is anticorrelated with optical emission, i.e. the polarisation could be diluted during the flares. 
However, the change in position angle during the flares suggests an additional emerging polarised component during the flares. 
We also find, based on our H$_\alpha$ spectroscopy and Doppler tomography, that there is indication for change in the accretion disc structure/emission during the flares, possibly due to a change in accretion flow. This, together with changing polarisation during the flares, could mark the existence of magnetic propeller mass ejection  process in the system. Furthermore, our analysis of flare profiles in both optical and NIR shows that NIR flares are at least as powerful as the optical ones and both can exhibit transition time scales less than 3 sec.
The optical/NIR flares therefore seem to originate from a separate, polarised transient component, which might be due to Thomson scattering from propeller ejected matter.  
\end{abstract}




\section{Introduction}

Radio millisecond pulsars (MSP) are fast rotating magnetic neutron stars (NS), which are thought
to be spun up during an evolutionary stage where they have been accreting matter (and angular momentum) from a mass-losing secondary star in a low mass X-ray binary (LMXB), i.e. the ``recycling process'' \citep{1982Natur.300..728A}. The final proof of this picture came in 2013 with the discovery of the first transitional millisecond pulsar (tMSP) IGR J18245--2452 \citep{2013Natur.501..517P}. 
We know now three similar systems -- PSR J1023+0038 \citep{2009Sci...324.1411A,2014ApJ...781L...3P}, XSS J12270--4859 \citep{2014MNRAS.441.1825B} and likely also 3FGL J1544.6--1125 \citep{2015ApJ...803L..27B} -- that exhibit transient behaviour between the accreting, LMXB state and non-accreting radio MSP state, thus enabling us to study the behaviour of the accretion stream/disc/structure when it is at the brink of existence.\footnote{3FGL J1544.6--1125 has only been observed in the accreting state.}

One of these systems, PSR J1023+0038, has remained in the accretion state since 2013 \citep{2013ATel.5514....1H,2014ApJ...790...39S,2014ApJ...781L...3P,2014ApJ...791...77T,2016ApJ...830..122J}.
The system consists of a NS and an evolved companion star in a 4.75h orbit with intermediate inclination \citep{2004MNRAS.351.1015W,2005AJ....130..759T}.
\citet{2012ApJ...756L..25D} give parallactic distance of $d = 1.37\pm0.04$~kpc, and NS mass of $1.71 \pm 0.16 \, M_\odot$ and \citet{2015ApJ...809...13D} found a flat radio spectrum, similar to other LMXBs in the same state; this is interpreted that there is a compact jet present.
The NS spin period of 1.69~ms has been detected both in the MSP state \citep{2009Sci...324.1411A, 2010ApJ...722...88A} and the LMXB state \citep{2016ApJ...830..122J}.
Also the optical spectrum currently shows double peaked hydrogen and helium lines, which is a clear signature of a newly formed accretion disc around the NS \citep{2013ATel.5514....1H,2014MNRAS.444.1783C,2015ApJ...806..148B}.

In the LMXB state, PSR J1023+0038 shows three flux modes; the high mode, the low mode and the flare mode, and X-ray pulsations are only seen in the most common high mode. 
That is, during flares and low mode the pulsations disappear \citep{2015ApJ...806..148B}.
The origin of these flux modes remains to be explained.
For example \citet{2014ApJ...785..131T} present a model for the increased $\gamma$-ray emission, which predicts that X-rays come from a shock outside the pulsar magnetosphere. 
However, such a scenario can only correspond to the non-pulsating low- and flare modes, since coherent X-ray pulsations can only come from the NS surface.

Further clues to the nature of the X-ray and optical flaring and mode switching were presented in a recent extensive study by \citet{2015ApJ...806..148B}, who analyzed XMM-Newton X-ray and optical/UV light curves from the Optical Monitor (OM), showing complex behaviour in between X-ray/UV vs. optical flaring.
Some X-ray flares had clear optical counterparts, while others did not.
More optical time series work was carried out by \citet{2015MNRAS.453.3461S}. They noted that the optical flares can have transition time scales of <20 sec and that short flares, lasting a couple of minutes, typically show a brightening of 0.1-0.5mag, whilst longer (5-60min) flares can achieve 1.0mag. 
\citet{2015MNRAS.453.3461S} propose a model for the flaring that involves ``clumpy accretion'' from a trapped inner accretion disc near the corotation radius (see also, \citealt{2012MNRAS.420..416D}).

\citet{2015ApJ...807...33P} present a model for propeller mode accretion in J1023, also suggested by \citet{2015MNRAS.453.3461S}, which seems to be supported  by the radio observations of \citet{2015ApJ...809...13D}.
Similarly, \citet{2016A&A...594A..31C} proposed a model where the rapid switching between the low- and high-modes corresponds to the NS alternating between the pulsar state and the propeller state, respectively.

Optical linear polarimetric observations of \citet{2016A&A...591A.101B} showed that PSR J1023+0038 has $\approx 1$\% level intrinsic polarisation degree in the V and R filters.
The origin of this signal is not fully clear, but it was attributed either to Thomson scattering in the accretion disc, or to synchrotron emission in the putative jet.
\citet{2016A&A...591A.101B} favoured the former idea based on the weak evidence of orbital modulation in the R filter polarimetric data.

In order to discern over these various scenarios, and in particular to study the optical flaring in more detail, we have performed a series of optical and NIR observations of PSR J1023+0038 between 2015-2017. Here we present the results of this campaign.

\begin{table*}
\caption{The observing log.}
\begin{center}
\begin{tabular}{|c|c|c|c|c|c|}
\hline
\hline
Telescope/instrument & Obs. mode & Date (MJD) & Time resolution & Duration (h) & Phase coverage \\   
\hline
\hline
SMARTS/ANDICAM &  V and H photometry & 57035-57049 & - & two week monitoring & \\
NOT/ALFOSC & White light photometry & 57069 & 2.7 sec &  1.1h & 0.71-0.94 \\
NOT/ALFOSC & White light photometry & 57070 & 2.7 sec &  0.8h & 0.58-0.75 \\ 
NOT/ALFOSC & White light photometry & 57071 & 2.7 sec &  1.1h & 0.94-1.18\\
ESO-NTT/SOFI & J band photometry & 57169 & 3.0 sec & 2.8 h & 0.73-1.36\\
NOT/ALFOSC & H$_\alpha$ spectra (grism \#17) & 57786 & 400 sec & 3.1h (29 spectra) & 0.74-1.05 \& 1.56-1.90  \\
 NOT/ALFOSC & H$_\alpha$ spectra (grism \#17) & 57787 & 400 sec & 7.1h (62 spectra) & 0.45-1.94 \\
  NOT/ALFOSC & H$_\alpha$ spectra (grism \#17) & 57788 & 400 sec & 6.5h (56 spectra) & 0.60-1.97 \\
   NOT/ALFOSC & H$_\alpha$ spectra (grism \#17) & 57789 & 400 sec & 4.4h (39 spectra) & 0.07-0.99 \\
NOT/ALFOSC & Polarimetry (filter \#92) & 57788 &  32 sec & 2.2 h & 0.57-1.03 \\
\hline
\hline
\end{tabular}
\end{center}
\label{default}
\end{table*}%

\begin{figure*}
	\includegraphics[width=\textwidth]{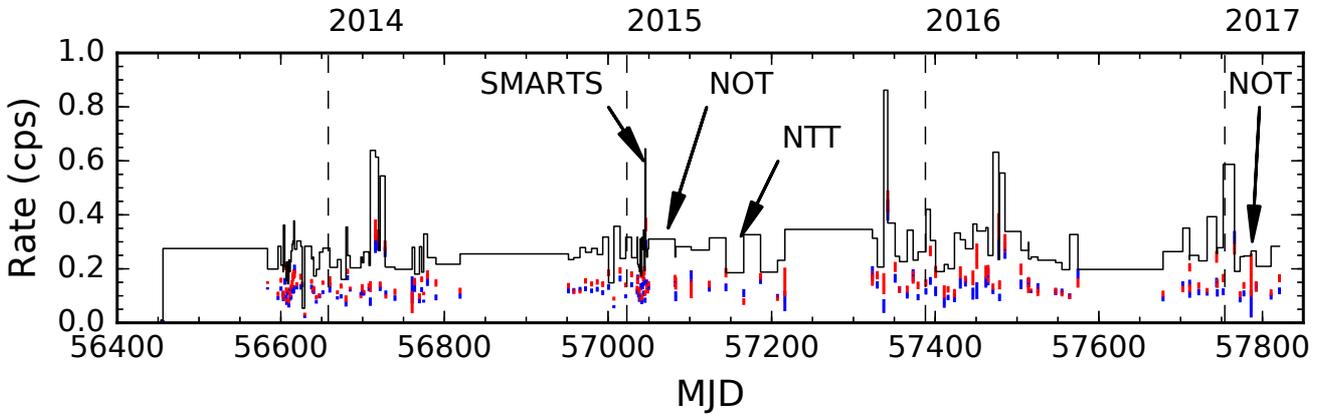}
    \caption{The epochs of our optical/NIR observations together with \textit{Swift}/XRT monitoring X-ray count rates (see Table 1). The red points mark the 0.3--1.5 keV count rate, the blue points the 1.5--10~keV count rate and the black line shows their sum.}
    \label{fig:testLC}
\end{figure*}


\section{Observations}

In this paper we cover optical photometry and polarimetry obtained in between 2015--2017  as well as NIR (J band) photometry from  2015. Finally we also present results from $H_{\alpha}$ spectroscopy carried out in Feb 2017.  The detailed log  of all observations  can be found in Table 1. Furthermore, the epochs of different datasets, together with \textit{Swift}/XRT \citep{2005SSRv..120..165B} X-ray count rates (obtained with the on-line XRT generator tool; \citealt{2009MNRAS.397.1177E}), are shown in Fig. 1. We have used the ``fiducial ephemeris'' of \citet{2016ApJ...830..122J} throughout the paper for computing the orbital phases.    

\begin{figure*}
	\includegraphics[width=16.0cm]{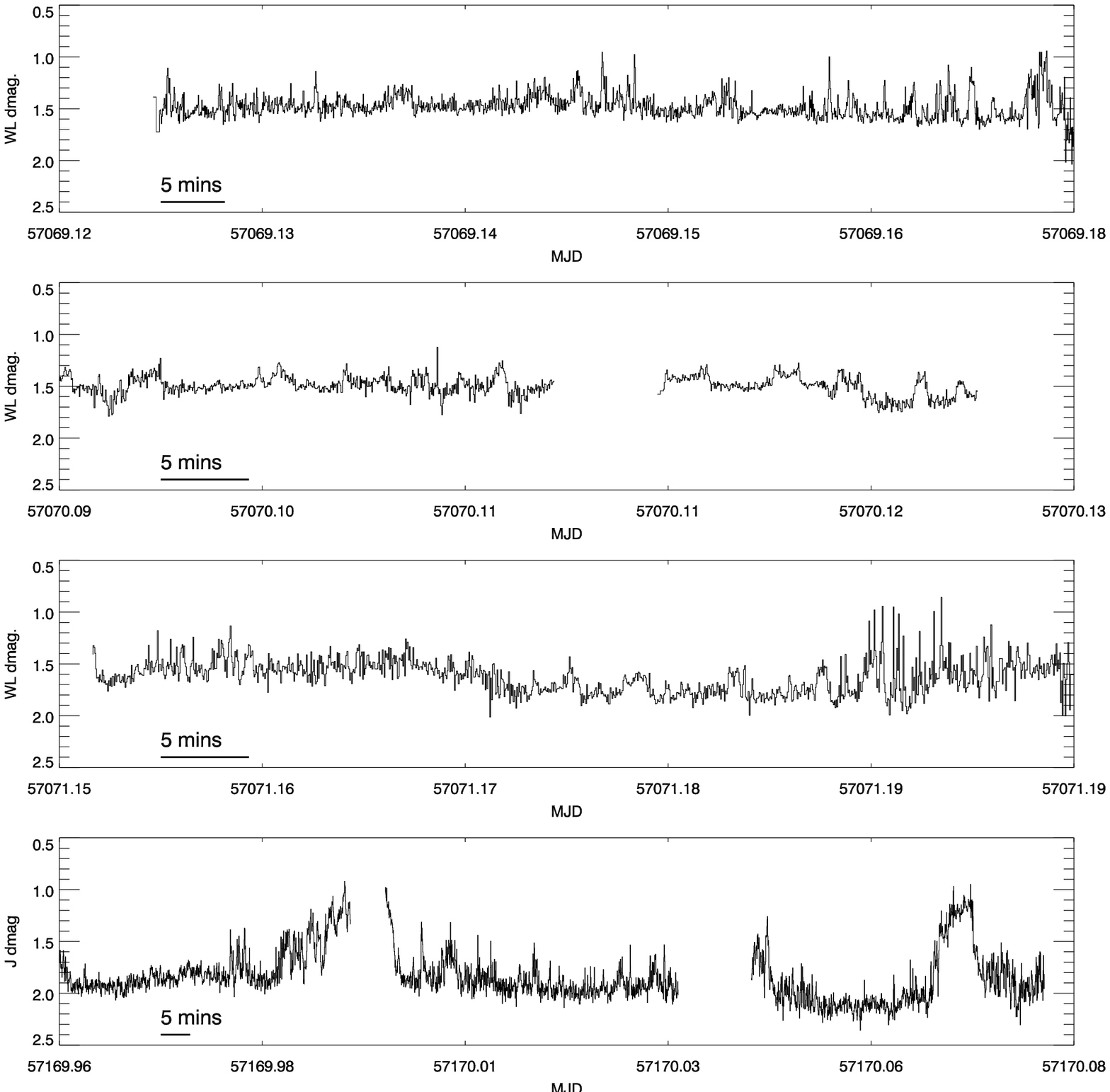}
    \caption{The fast photometry light curves of J1023. The top three panels show white light optical data from NOT and the the bottom plot shows the ESO NTT/SOFI J band photometry obtained 3.5 months later. The time resolution is 3 sec in all data. The horizontal bars mark 5 min length scale.}
    \label{fig:smarts_lc}
\end{figure*}

\subsection{Optical and NIR photometry}

Optical photometry was carried out on the 2.5m Nordic Optical Telescope (NOT), La Palma in Feb. 2015 using ALFOSC, a multimode imager/spectrograph/polarimeter. This consisted of fast photometry (1sec exposure times, 2.7sec effective time resolution, cf. Table 1) in white light. The data  were bias subtracted and flatfielded in the usual manner using IDL. We used a small CCD window, that yielded a field that also contained a comparison star located at R.A.(2000)=$ 10^h23^m43.31^s$, Dec(2000)=$+00^o38'19.2"$, which has V=14.86 \citep{2005AJ....130..759T} and H=13.16 in the 2MASS catalogue \citep{2006AJ....131.1163S}.
The differential white light magnitudes quoted are relative to this source. The light curves are shown in Fig 2.

\begin{figure}
	\includegraphics[width=9cm]{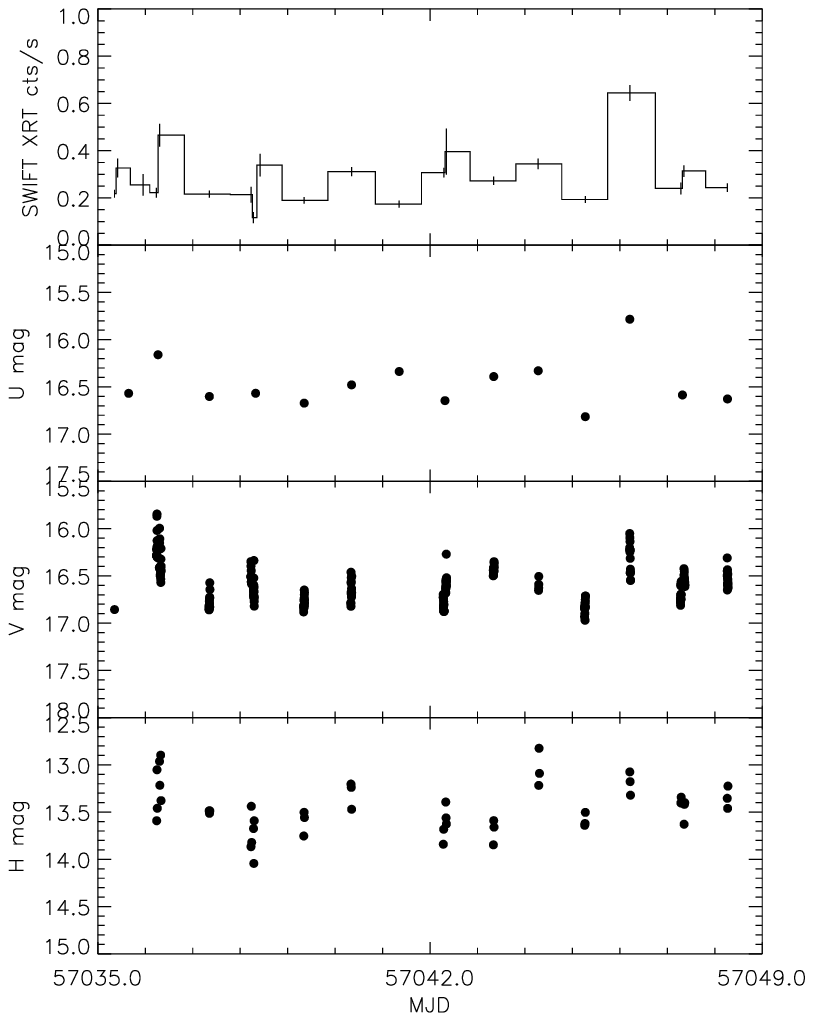}
    \caption{\textit{Swift}/XRT + UVOT U band and SMARTS/ANDICAM V + H band monitoring of J1023 in January 2015.}
    \label{fig:smarts_lc}
\end{figure}

We also obtained some fast NIR photometry in J band using NTT/SOFI in May 2015 as a part of project to look for orbital periods of MSPs. Observations were carried out for about 2.5--3~h on two consecutive nights, but the data from the second night are very noisy due to highly variable observing conditions. We thus only present data from the first night. The observations were obtained in J band, where we could achieve the best possible time resolution by still being able to use the stare mode. We employed a 160x600 pixel subwindow and the exposure time was 3sec yielding also an effective time resolution of 3~sec for the datacubes. The same comparison star as in NOT observations was included in the frame. The resulting datacubes were bias subtracted and flatfielded in the same way as optical data, since we used the stare mode. The resulting differential J band light curve is shown in Fig. 2. Some more simultaneous optical and NIR (V+H bands) monitoring data were obtained in Jan/Feb 2015 over a period of two weeks using SMARTS/ANDICAM operated by the SMARTS Consortium (Fig. 3). These data were bias subtracted and flatfielded (V band). The dithered H band data were darksubtracted and flatfielded, after which a single image was created from a sample of 7 dithered pointings (following a removal of median sky frame from the same set of 7 pointings).

\subsection{Optical polarimetry}

Time series optical linear polarimetry was obtained using the NOT in Feb. 2017. These observations cover only a single run of 2.2~hrs with 32~sec time resolution. The data were taken with ALFOSC equipped with the WeDoWo \citep{1997A&AS..123..589O}, a Wedged Double Wollaston polarimeter module, that is capable of measuring a polarimetric datapoint from a single exposure by splitting the light beam into four sub-beams with 0, 45, 90 and 135 degree polarisation angles. The same nearby comparison star, as mentioned above, was included in the wide 10" slit for polarimetry. This provided simultaneous monitoring of the polarimetric conditions. The images were bias subtracted in a normal manner, after which they were flatfielded and transmission corrected for differences in different light beams using dome flats taken with the same rotator angle and with a rotator angle with 90 degrees offset. As the observations were taken using a non-standard wide sky contrast filter (ALFOSC filter \#92, appx. 4000-7000\AA), there are no suitable polarimetric standards available. However, we did correct the position angle using observations of a high polarisation standard BD+64 106, for which the position angle is known and does not vary significantly with colour over the wavelength range in question \citep{1992AJ....104.1563S}. 

The polarimetric data were taken during the last night of the observing run in Feb. 2017.
Given the characteristics of the WeDoWo polarimeter module, the correction for zero offset for polarisation would have required a series of zero polarisation standard observations with different instrument rotator angles.
However, given the time limitations and the exploratory nature of this 2.2~h run, zero polarisation standard observations were inadequate to fully account for field dependent instrumental polarisation, and thus the absolute values of the degree of polarisation are not calibrated. 
We remind, however, that this only introduces uncertainty in the absolute level of polarisation, not the changes in polarisation during our time series. 
The inclusion of the comparison star in the same slit does provide us the necessary control over non-astrophysical changes in polarisation, which was the main driver for these observations. 
The resulting polarisation curve for the source is shown in Fig. 4 and the individual polarisation measurements, both for the source and the comparison star, are shown in Stokes (Q/I, U/I)-plane in Fig. 6. 


\subsection{H$_{\alpha}$ spectroscopy}

We also obtained H$_\alpha$ spectroscopy of J1023+0038 during the Feb 2017 NOT run  (see Table 1 for detailed timing and orbital phase information of the spectra). These observations were also carried out with ALFOSC using a grism \#17, together with 2x on-chip binning both in spatial and wavelength direction for increased S/N. This, together with a 0.9" slit, yielded a spectral resolution of R=6000 over a wavelength range of 6350-6850\AA. We used 400~sec exposure times. The slit was oriented along the parallactic angle, apart for the last night, when a comparison star was included in the slit to better monitor the flaring in the source.
Standard CCD reductions were applied and the spectra were extracted using the OPTSPECXTR IDL-routines by J. Harrington that employ the optimal spectral extraction algorithm from \citep{1986PASP...98..609H}. Each individual spectrum was finally wavelength calibrated using the sky lines around H$_\alpha$, as this was deemed more accurate than using arc lamp spectra obtained about once in two hours. Given that our aim was to solely study the H$_{\alpha}$ profile and its changes (and the narrow slit), no spectrophotometric standards were observed. 

\begin{figure}
	\includegraphics[width=8.5cm]{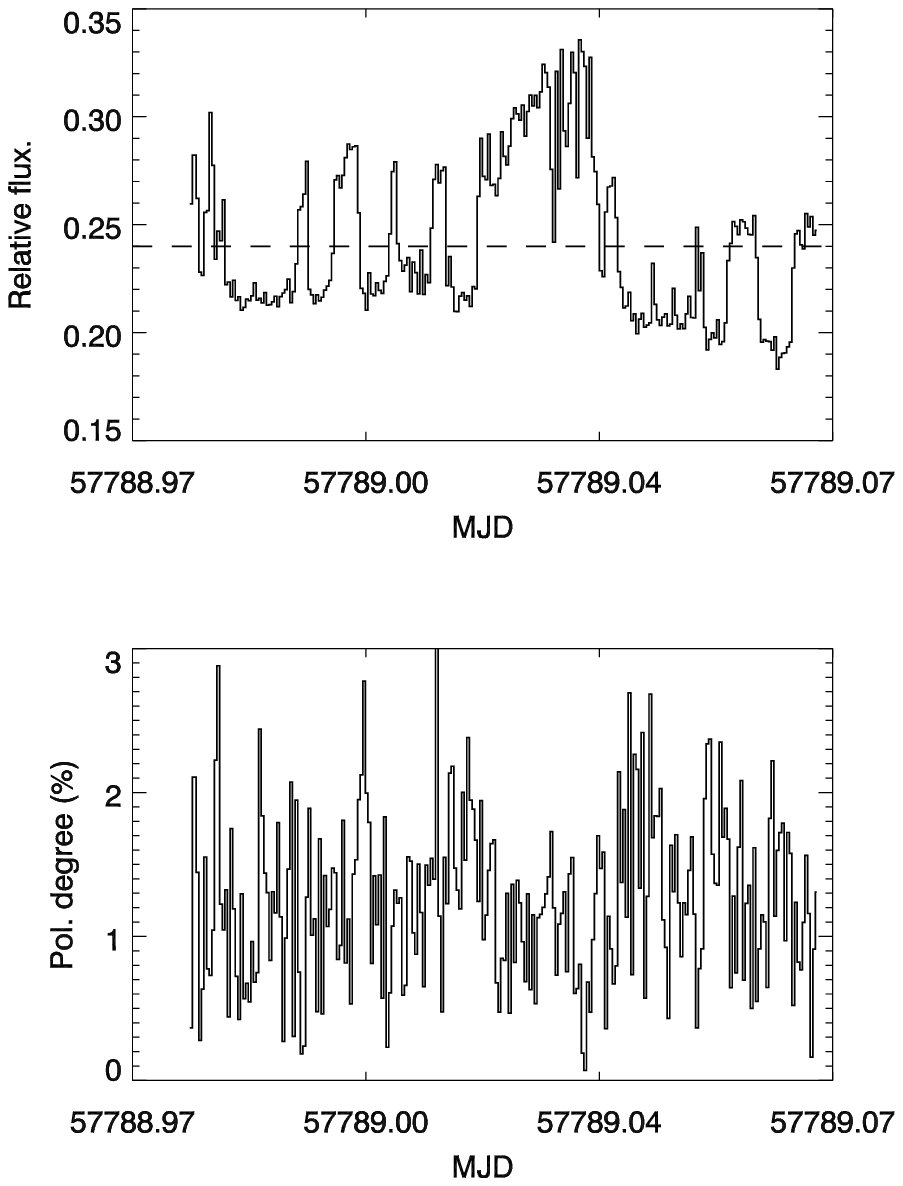}
    \caption{Simultaneous photopolarimetry of J1023+0038 in wide (4000-7000\AA) ``sky contrast'' band. The differential light curve (top) and degree of linear polarisation (bottom) are shown. The dashed horizontal line (top) indicates the division used for identifying flaring/non-flaring states.}
    \label{fig:smarts_lc}
\end{figure}
\begin{figure}
	\includegraphics[width=8.5cm]{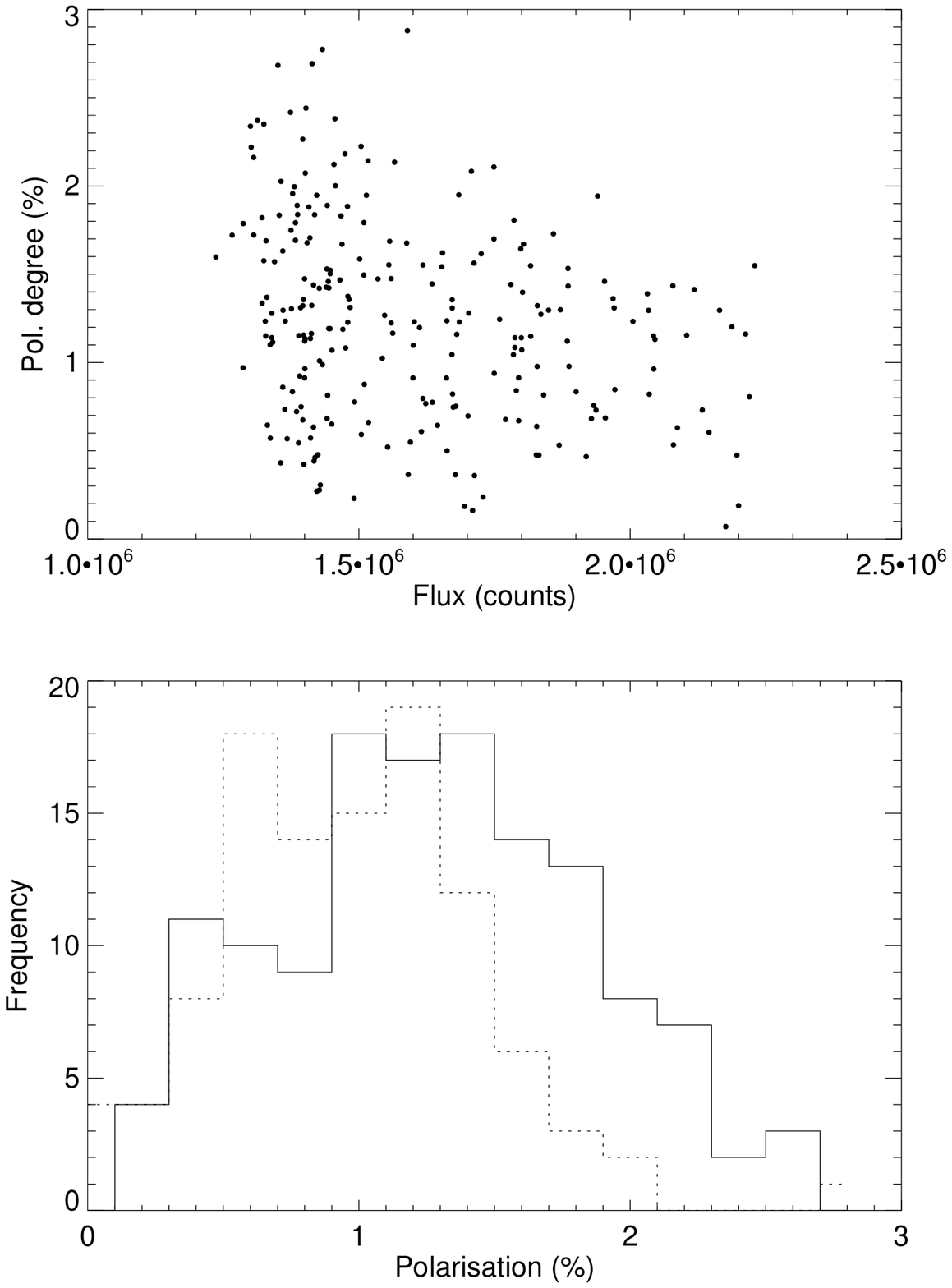}
    \caption{The anticorrelation between source flux and polarisation degree (top) and the histograms of polarisation measurements for the "non-flaring" (solid) and flaring (dotted) states (see Fig 3. and text for the division criterium).}
    \label{fig:smarts_lc}
\end{figure}

\section{Results and Discussion}

\subsection{Optical and NIR fast photometry}
Our four fast  photometry light curves are shown in Figs 2 (and partially zoomed in Fig. 7). The top three  panels of Fig. 2 show the optical (white light) light curves obtained at the NOT using ALFOSC in Feb 2015, whilst the bottom panel shows our single J band light curve obtained at the ESO NTT /SOFI 3.5 months later in May 2015. The time resolution of each light curve is 3 sec. We indicate different time scales of different light curves by horizontal line marking a 5 min time span. 

All our optical light curves are dominated by short  (less than  3 min) flares, most clearly visible during the second night (MJD=57070, second panel in Fig 2.), when the  S/N was best.  There appear to be at least two types of flares, some of which last longer (several minutes) and  are characterised by flat tops and  emission peaking  right after the rise and just before the decline. Other flares last a shorter time (appx. 1 min or less) and seem to lack the flat tops. We don't observe any larger flares, like the one lasting 28 min in our polarimetry, during these observations. The flare amplitudes are typically of the order of 0.2-0.25mag, similar to the smaller flares during the polarimetry run, whilst the large flare during the polarimetry reached 0.4mag amplitude. These values are similar to what's been reported by \citet{2015MNRAS.453.3461S} and \citet{2015ApJ...806..148B}.  

We also present the first ever NIR (J band) fast photometry of the system in flaring state. Our J band data, although covering only a single 2.8h run and not backed up by simultaneous optical data, reveals that at least in the J band the flares can exceed 1 mag, as demonstrated by the single 8.5min flare that was totally covered by J band data (Figs 2 \& 7.). Unfortunately two other NIR flares partly coincided with data gaps, as seen in the bottom panel of Fig 2., but especially the first event showed, judging by the ingress and egress, likely more than 1 mag increase in the source brightness. The profile of the totally observed NIR flare resembles that of the large optical flare observed during our polarimetry (Fig 4.). 

Both ingresses and egresses of the flares appear equally rapid. In some cases they are not even resolved in our data that has a time resolution of 3 sec. For instance, there is a $\sim$0.3 mag drop in the J band flare that takes less than 3 sec. Similar timescales can be observed in much weaker optical flares (Fig. 7). We find that most flares are "rectangular shaped" and appear to have "horns" i.e. they show maximum brightness right after the ingress and just before egress. It is not possible to conclude, based on our data, whether this marks the intrinsic emission profile during the flare, or whether there is some extra absorption during the flare that causes the mid-flare emission to be attenuated. It is intriguing to note though that our single J band flare does not show such structure. However, as noted earlier, the profile of this NIR flare resembles more the long lasting (28 min) large optical flare in our polarimetry and its length (8.5 min) is much longer than the rectangular optical flares seen here (1.5-2.5 min). Thus the single J band flare might not be comparable in terms of shape to the optical flares presented here.    

The quasi-simultaneous \textit{Swift} and  SMARTS/ ANDICAM monitoring of the source in X-rays and U, V and H bands over a period of two weeks in January 2015 (Fig 3.) revealed a single strong X-ray flare, which was also detected
in U, V and H bands (the epoch of highest \textit{Swift} XRT flux in Fig 3. at MJD=57046). The tripling of X-ray flux (from 0.2 to 0.6 cts/s) after MJD=57046 was reflected by a 1 mag brightening in U band and 0.5 and 0.3-0.4mag in V and H bands, respectively, confirming that the NIR, optical and X-ray flaring are produced by the same phenomenon.
Finally, there is no clear indication of the sinusoidal orbital modulation in our light curves (as reported by \citet{2014MNRAS.444.1783C} and \citet{2015MNRAS.453.3461S}. However, all of our fast photometry light curves only cover a fraction of the orbital period and such modulation might then easily go undetected. Our 2.2h photopolarimetric run that covers orbital phases 0.57-1.03 (Fig 4.) does however show some evidence for possible underlying modulation.

\subsection{Optical polarimetry}

As noted earlier, the source was observed in a  polarimetric mode using ALFOSC, equipped with WeDoWo ``one-shot''  polarimeter module in February 2017. We obtained only a single polarimetric run, lasting about 2.2h, as the main aim of our observing run was to study the possible changes in the accretion  disc structure in flaring/non-flaring states using Doppler tomography. However, even during this short pilot run, we  were able to see several flaring events, including one large flare lasting about 28 minutes. The resulting light curve (relative to the comparison star in the wide 10" slit) and linear polarisation degree curve are shown in Fig 4.  In addition to the  number of flares, there also appears  to be striking anticorrelation between flux and degree of polarisation, especially visible during the larger flare in the middle of the run. We carried out a series of tests to verify the significance of the apparent anti-correlation. Firstly, Fig. 5 shows the flux vs. degree of polarisation (top). We computed a rank correlation test (Kendall's $\tau$), which returned a chance probability of 1.2$\times 10^{-5}$ for the correlation to appear at random. Secondly, we divided all the data in two groups based on whether the source was flaring or not during the observation. This was carried out using flux cutoff at 0.24 level in Fig 4 (top). 
This particular cutoff value was chosen so that it optimised the separation of flaring and non-flaring flux levels during the time series.
This resulted in two samples with 134 observations in non-flaring and 107 observations in flaring state. Next we proceeded to compare polarisation measurements obtained for these two samples using various tests. The key idea was to test at what significance level do the two distributions of polarisation measurements differ, if at all. We started by computing a two sample Kolmogorov-Smirnov test, which resulted in a chance probability (p value) of 2.2$\times 10^{-4}$. Next we proceeded to compare the samples with more rigorous tests like the Mann-Whitney U-test and the K-sample Anderson-Darling test for two samples, which are more sensitive to the changes in the median value than the K-S test and which produced p-values of 1.7$\times 10^{-5}$ and 1.6$\times 10^{-5}$ respectively. We also computed the same tests for the comparison star using the same two samples as for the source. This produced chance probabilities of 0.44, 0.70 and 0.54 for the K-S, Mann-Whitney and Anderson-Darling tests respectively.  It does appear that the detected anti-correlation is indeed real in terms of statistical analysis. We have also considered the possibility of flux dependent instrumental polarisation. This would require significant spatial variations in the linearity of the CCD-chip, since in wedged double Wollaston configuration there are no moving optical parts (during our time series) and the polarisation degree is computed from each image individually. Furthermore, autoguiding was used during the run restricting any image shift during the run. We are not aware of such non-linearities in the CCD detector. It is thus very unlikely that the observed anticorrelation is of instrumental origin.

\begin{figure}
	\includegraphics[width=9cm]{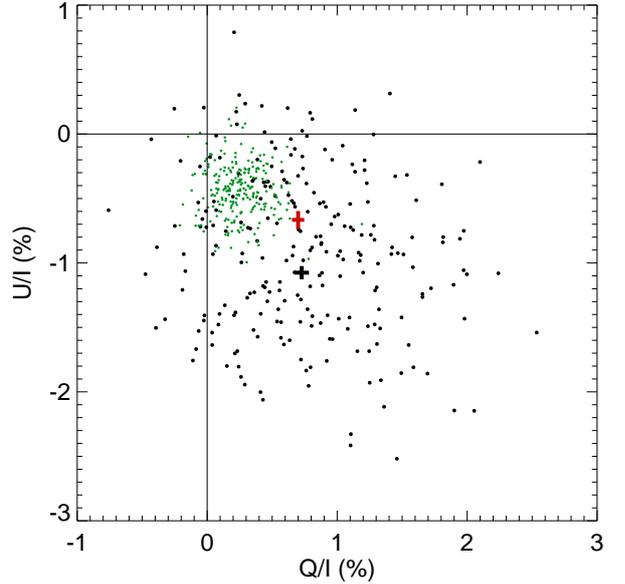}
    \caption{The polarisation data in Stokes Q/I vs. U/I plane. The black (larger) dots show J1023 and green (smaller) dots the comparison star data obtained simultaneously. The red (top) and black (bottom) ``crosses'' (actually X- and Y-error bars) mark the mean values (and their 1$\sigma$ errors) for the J1023 polarisation in flaring and non-flaring states respectively.}
    \label{fig:smarts_lc}
\end{figure}

\begin{figure*}
	\includegraphics[width=19.0cm]{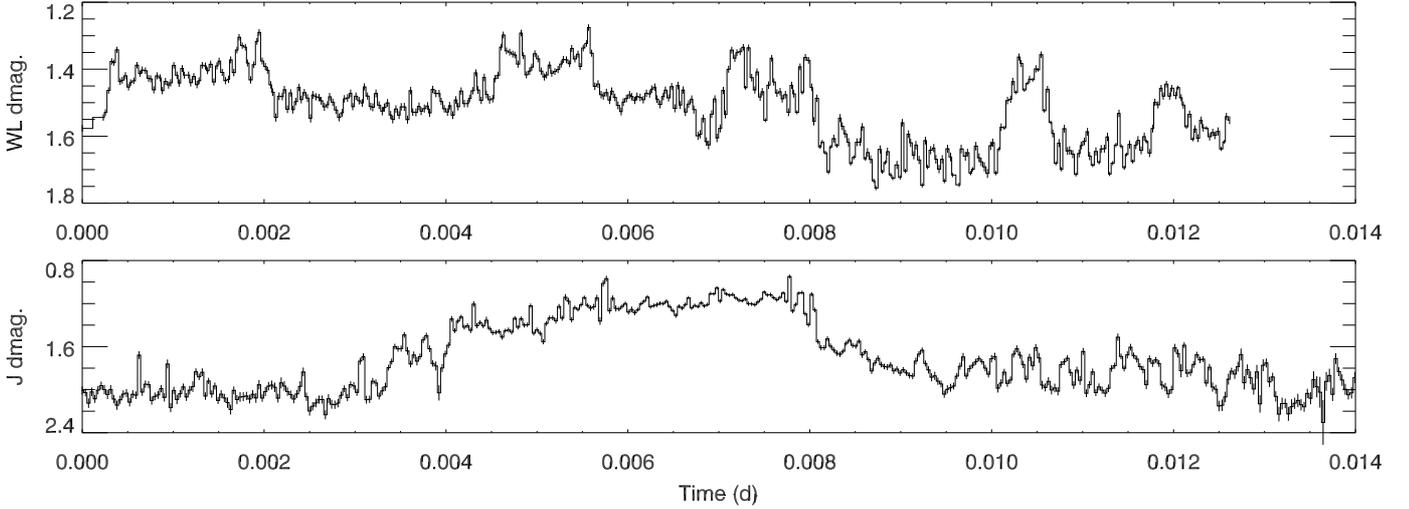}
    \caption{The zoomed in fast photometry light curves of J1023. The data are from 16th Feb (optical, top) and 28th May (J band, bottom) 2015 and show the latter part of the second and fourth panels of Fig. 2. Note that even though the time axes are comparable, the magnitude scales are very different.}
    \label{fig:smarts_lc}
\end{figure*}

\begin{figure}
	\includegraphics[width=9cm]{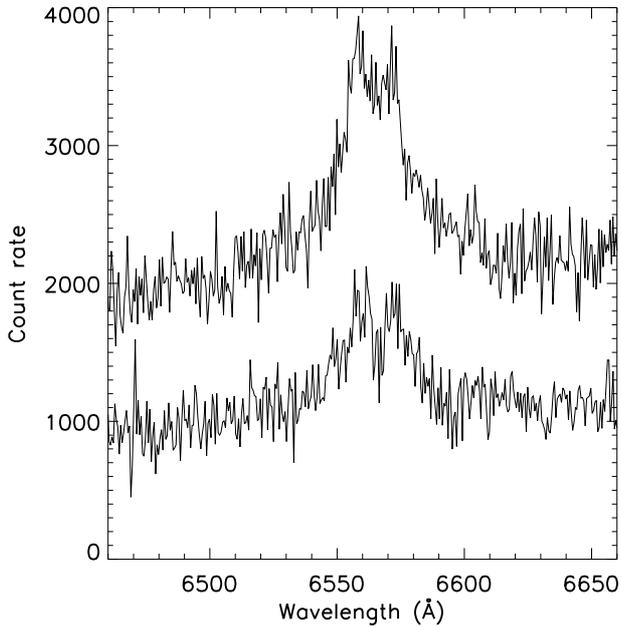}
    \caption{Sample individual spectra taken during flaring (top) and non-flaring (bottom) states. These were taken on MJD=57788 and correspond to the orbital phases 0.46 (flaring) and 0.25 (non-flaring). Note the factor of two difference in the count rate.}
    \label{fig:smarts_lc}
\end{figure}

It is interesting to note that the change in system polarisation during a flare immediatedly implies that at least some of the observed polarisation is intrinsic to the source. The nearby comparison star, observed simultaneuosly in the same slit, also shows polarisation, but at a lower level. We have plotted the polarisation data in Stokes Q/I vs. U/I plane in Fig 6. The J1023 datapoints are shown with black dots, whilst those of the comparison star are shown in green. Both sources appear to show roughly the same position angle, suggesting that at least some interstellar polarisation is present. More data in different colours is needed to measure the amount of interstallar polarisation. The source is about 1.5mag fainter then the comparison star in our wide band data. \citep{2005AJ....130..759T} found it to be 2.6mag fainter than the same comparison star in their V band data. This was prior to the accretion state started though. The smaller degree of polarisation associated with the comparison star is likely due to its smaller distance and thus less interstellar extinction. 

The black and red ``crosses'' in Fig. 6 mark the median values for the polarisation of J1023 in its flaring (red) and non-flaring (black) state. The arms of the crosses actually denote the 1$\sigma$ error of the mean in Q/I and U/I. It is again interesting to note that the change in polarisation during the flares happens only in U/I direction. There is no significant change in Q/I. This potentially has implications on the underlying physics, since a simple increase of non-polarised flux during the flare should have similar diluting effect on both the Q/I and U/I components. However, as total polarisation is a vector sum of subcomponents such behaviour, as observed here, tends to suggest that the change in polarisation is rather caused by an addition of another polarised emission component, that happens to be ``aligned'' along the U/I axis. Such effect could be produced by extra scattering matter emerging during the flares, which could be the case if some of the accretion flow would be ejected from the system. This could be driven by the magnetic propeller process, which has been suggested to operate in the system \citep{2015ApJ...807...33P, 2015MNRAS.453.3461S}. 

\begin{figure*}
	\includegraphics[width=15cm]{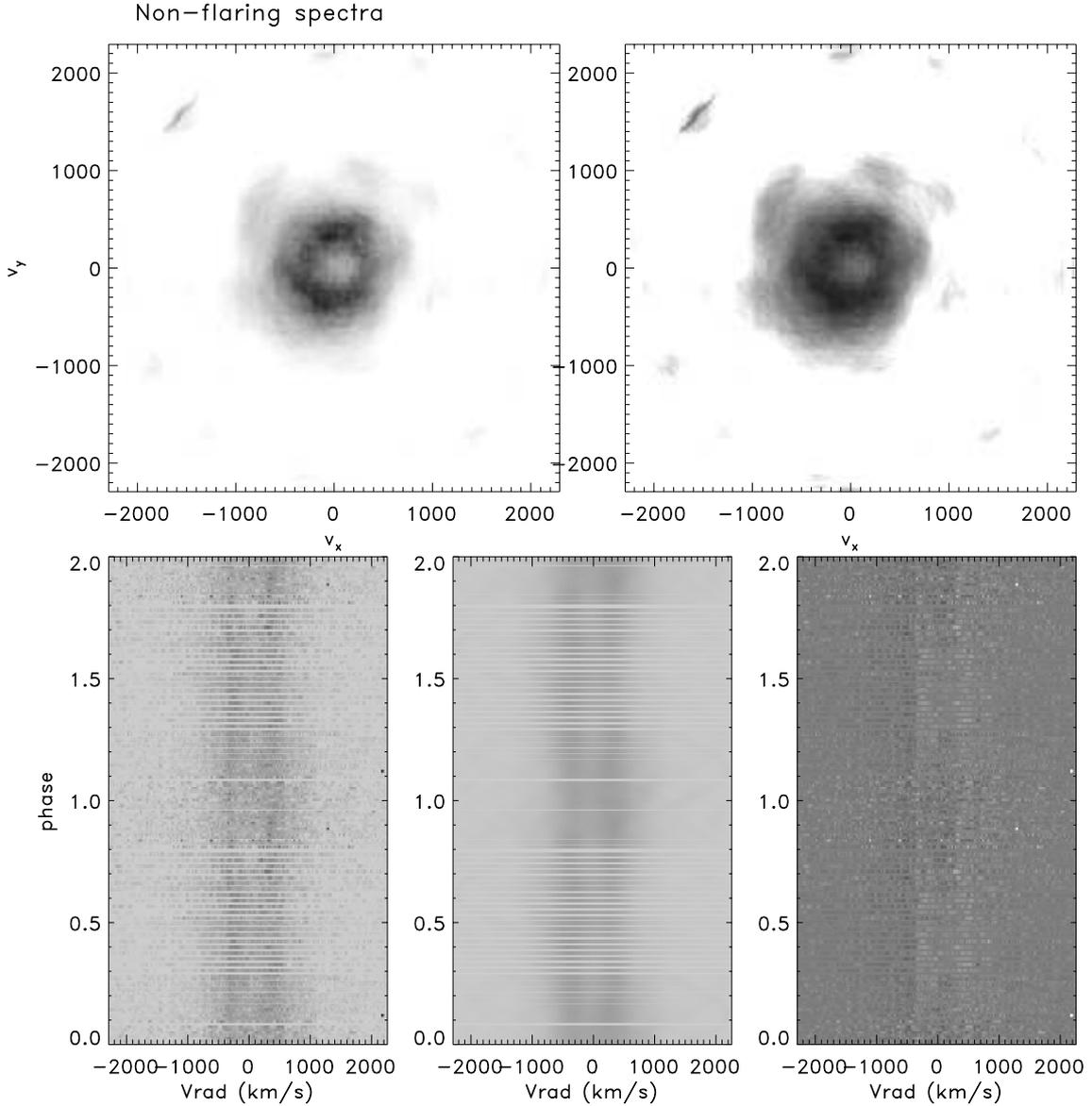}
    \caption{Doppler tomogram of H$_\alpha$ line obtained using spectra in non-flaring state. Linear mapping (top left), log-mapping (top right), trailed spectrogram (lower left), the model (lower middle) and residuals (lower right) are presented. The emission is fairly axisymmetric indicating a "normal" accretion disc.}
    \label{fig:testLC}
\end{figure*}
\begin{figure*}
	\includegraphics[width=15cm]{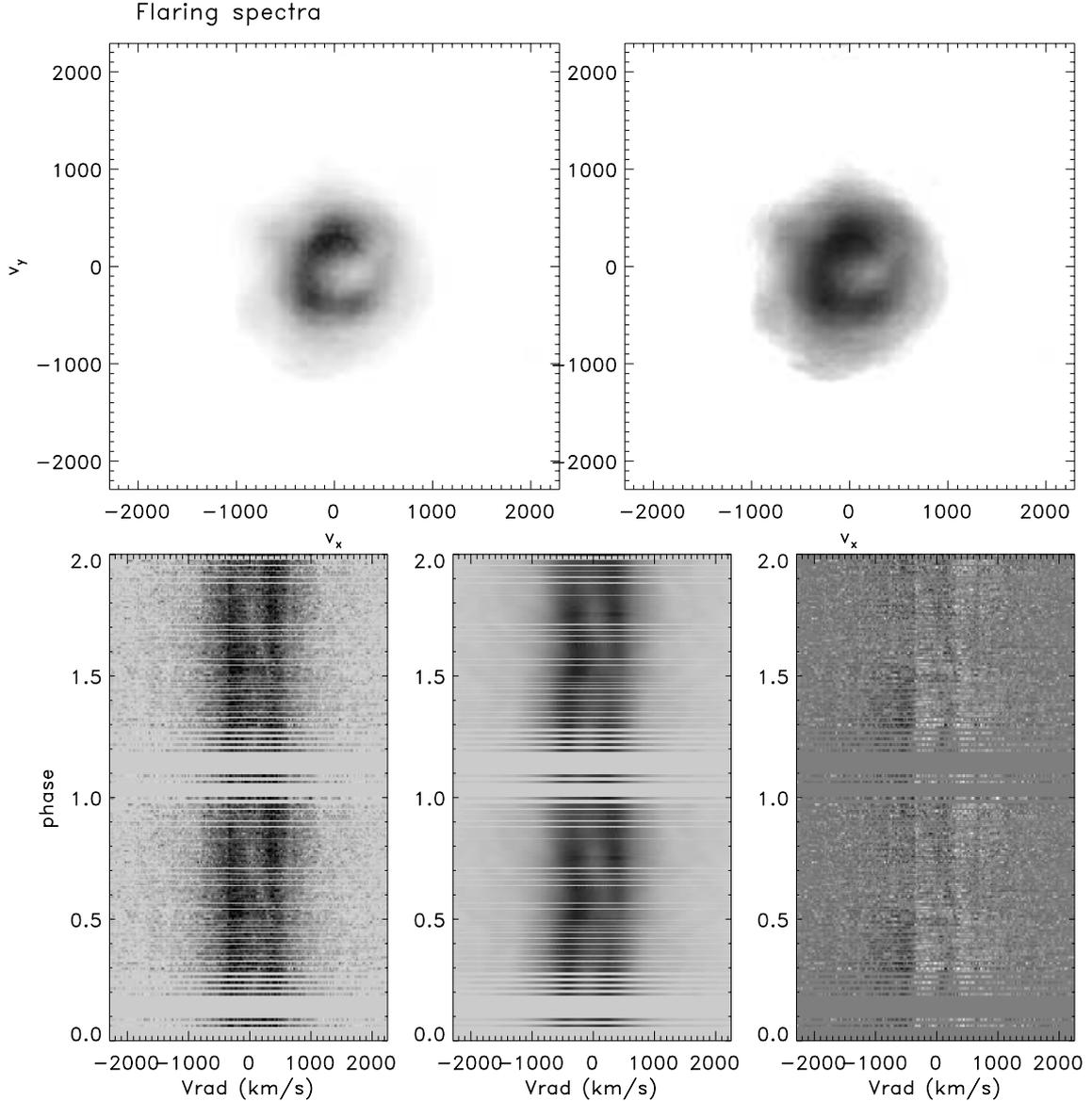}
    \caption{Doppler tomogram of H$_\alpha$ line obtained using spectra in flaring state. Linear mapping (top left), log-mapping (top right), trailed spectrogram (lower left), model (lower middle) and residuals (lower right) are presented. The emission is somewhat non-axisymmetric, with less emission coming from near maximum $V_x$ and $V_y$ near 0.0. This marks the area of disc closest to the observer at phase 0.25.}
    \label{fig:testLC}
\end{figure*}

\begin{figure*}
	\includegraphics[width=18cm]{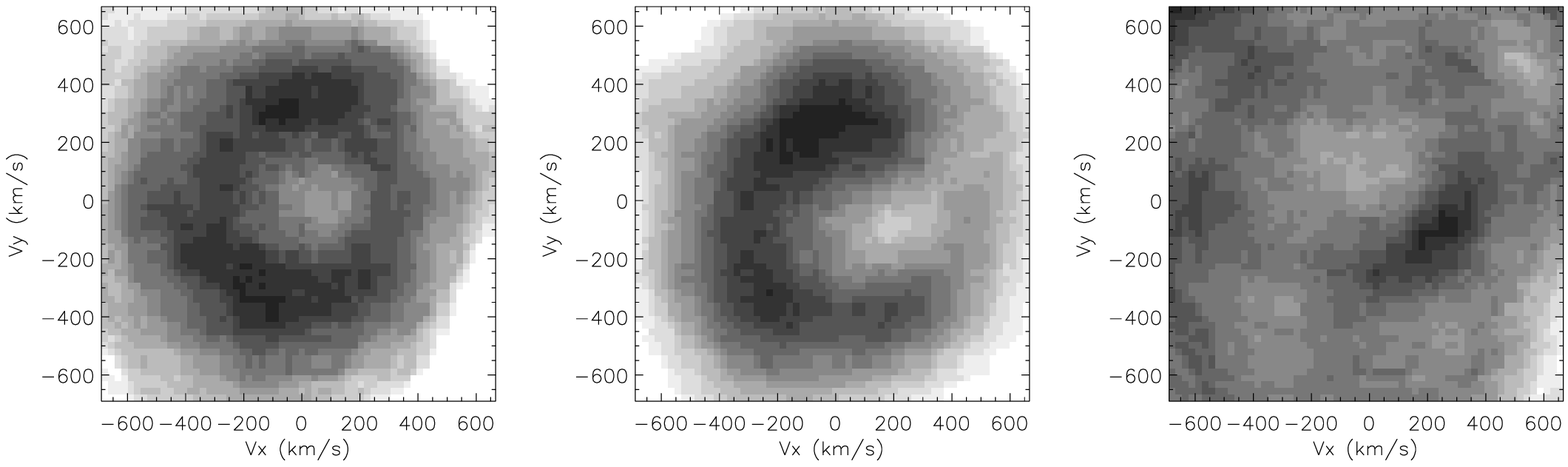}
    \caption{The non-flaring (left) and flaring (centre) Doppler tomograms of H$_\alpha$ line compared. The tomogram on the right shows the residual map.}
    \label{fig:testLC}
\end{figure*}

\subsection{H$_\alpha$ spectroscopy and Doppler tomography}

The main aim of the February 2017 observing run at the NOT was to study the accretion disc structure, compared to ``normal'' constantly disc-accreting LMXBs. As the source was flaring frequently, we managed to build up phase resolved spectra during both, the non-flaring and flaring states. The division of spectra into these classes was based on slit fluxes during the first three nights, when a parallactic angle was used. On the last night the comparision star was included in the slit. As with the polarimetric data earlier, the slit fluxes showed clear difference in between flaring  and non-flaring states (see Fig. 8). This was true during each of the nights, independent of the inclusion of the comparison star in the slit or not. Even if the observing conditions were photometric during each night and the division is based on the bimodality of the slit count rates it should be taken with some caution. We show a typical flaring (top) and non-flaring (bottom) individual spectrum in Fig.~8 using slit count rates. 

Having  divided the spectra into ``flaring'' and ``non-flaring'' samples, we proceeded to carry out Doppler tomography of the accretion disc based on the H$_\alpha$ line. This was performed using the DOPMAP code of \citet{1998astro.ph..6141S} separately for the flaring and non-flaring spectra. The results, along with the trailed spectra, are shown in Figs 9 and 10. In both cases  a clear double peaked line, suggesting the presence of an accretion disc, is observed. In addition, the relative strength of the blue and red peaks varies in flaring spectra over the orbital period. 

This results in Doppler tomograms of the flaring spectra, where the emission distribution is not entirely axisymmetric, as there is clear lack of emission  at around maximum $V_x$ (where $V_y$ is around zero or slightly negative), or the ``third quarter'' (i.e. $V_x$ >0 and $V_y$ <0) in the Doppler tomogram in general (Fig. 11). Assuming an accretion disc like (Keplerian) velocity field, this corresponds to the sector of the disc that is closest to the observer in between phases 0.25-0.5. Such lack of emission in flaring state would be compatible with part of the matter being ejected from the system by a propeller mechanism (see \citealt{1997MNRAS.286..436W} for the predicted Doppler tomogram of the propeller model).

As mentioned earlier, the propeller model has been previously suggested by Papitto \& Torres (2015), Shahbaz et al. (2015) and Deller et al. (2015). Whilst our Doppler tomograms alone cannot confirm the existence of propeller process in J1023, our results demonstrate that there are definite changes in the line emission from the accretion flow during the flares. Furthermore, our polarimetric results seem to support this interpretation. 

\section{Conclusions}

Our optical and NIR observations of J1023+0038, including the first NIR fast photometry of the system, reveal stronger flaring in J band than is observed in the optical. Both the optical and NIR flares show that timescales shorter than 3 sec can be observed during the flare transitions. 
Our optical polarimetry of J1023+0038 confirms that the system shows intrinsic optical linear polarisation, which is observed to vary during a large optical flare.
The variation could be due to an increase in non-polarised emission, but as it is only observed in one Stokes parameter, it is more likely to result from an additional polarised emission during the flare.  Such emission could be due to Thomson scattering from matter ejected by the propeller mechanism during the flare. Alternatively, it could arise from the emerging synchrotron emission during the flaring. Very recently (since the completion of this work) the source was detected as an optical pulsar \citet{2017arXiv170901946A} lending support to the synchrotron origin of the polarised emission.  

Finally, our H$\alpha$ spectroscopy/Doppler tomography indicates that there are changes in the accretion disc emission distribution in between the ``flaring'' and ``non-flaring'' states. The similarity of our resulting ``flaring'' state Doppler tomograms with the ones expected from the magnetic propeller \citet{1997MNRAS.286..436W} suggests that the propeller effect might occur during flaring. This has also been suggested earlier by \citet{2015MNRAS.453.3461S}.
More time series polarimetry, fast photometry, preferably  simultaneously in optical and NIR, as well as spectroscopy are clearly needed to resolve the origin of the flares and the role of possible magnetic propeller in the system.    

\section*{Acknowledgements}

JJEK acknowledges support from the V\"ais\"al\"a Foundation and from the Academy of Finland grant 295114. 
Based on observations made with the Nordic Optical Telescope, operated by the Nordic Optical Telescope Scientific Association at the Observatorio del Roque de los Muchachos, La Palma, Spain, of the Instituto de Astrofisica de Canarias.
The data presented here were obtained in part with ALFOSC, which is provided by the Instituto de Astrofisica de Andalucia (IAA) under a joint agreement with the University of Copenhagen and NOTSA.
Based partly on observations collected at the European Organisation for Astronomical Research in the Southern Hemisphere under ESO programme 095.C-0185(B). We acknowledge the use of DOPMAP package by H. Spruit and the OPTSPECEXTR IDL-routines by J. Harrington. Finally, we wish to thank the anonymous referee for the constructive comments that helped us to improve the paper.




\bibliographystyle{mnras}
\bibliography{J1023_refs} 




\bsp	
\label{lastpage}
\end{document}